\documentstyle[prl,multicol,aps]{revtex}
\begin{document}
\title{Linear Chain of Coupled Quantum Dots}
\author{Kicheon Kang}
\address{Department of Physics, Korea University, Seoul 136-701, Korea}
\author{Min-Chul Cha}
\address{Department of Physics, Hanyang University, Kyunggi-do,
         Ansan 425-791, Korea}
\author{S.-R. Eric Yang}
\address{Department of Physics, Korea University, Seoul 136-701, Korea}
\maketitle
\draft
\begin{abstract}
A linearly coupled chain of spin-polarized quantum dots is  
investigated  under the condition that the number of 
electrons is equal to or less than the number of the dots.
The  chemical potential of the system, $\mu_{N}=E(N)-E(N-1)$, 
satisfies, $(\mu_{N}+\mu_{N_{\ell}+2-N)}/2 \approx V+2t$
($N$, $N_{\ell}$, $V$, $E(N)$ and $t$ are the number of 
electrons, the number of dots, and the strength of nearest 
neighbor electron-electron interactions, the total groundstate 
energy and the hopping integral  between two adjacent dots). 
This property will be reflected in the spacing between the 
conductance peaks.  The electron density structures  are 
determined using a quantum Monte Carlo method.  As the number 
of electrons is varied several correlated structures are found 
that are commensurate/incommensurate with the periodic dot system.
Hartree-Fock theory fails to predict the correct electronic
structures of this system because several nearly degenerate
solutions exist.
\end{abstract}
\thispagestyle{empty}
\pacs{PACS numbers: 73.20.Dx, 73.20.Mf}
 
\begin{multicols}{2}
Recent advances in nano-fabrication techniques have made it possible
to make quantum dots \cite{K,Kas}.  
These structures have several 
similarities to atoms and are called artificial 
atoms. 
The physical parameters of these systems may be controlled easily:
the electron density can be varied significantly 
by the substrate voltage $V_{g}$ and 
the range of electron-electron interactions
can be controlled by changing the distance to the metallic layer.
When these
artificial atoms are coupled structures similar 
to a molecule may arise.
Recently numerous groups have started to
investigate whether coupled two-dot systems really have molecular
properties.  Properties of a two-dot
system have been studied both experimentally\cite{Kou,Wau,Bli,Vaa,Hof}
and theoretically\cite{Mat,Gol}.
When many dots are coupled an artificial one-dimensional crystal
may be created.  It is possible 
to couple up to ten or more dots together
since the size of the total system can be made smaller than the 
phase coherence length.
Such a system has energy bands separated by energy gaps, and
transport properties of such a periodic crystal have been investigated
experimentally \cite{Kou2}.

When the length of such a periodic crystal is infinite the system 
exhibts many intriguing properties.  Depending of the values of the physical
parameters the system can have properties of a Luttinger liquid 
or a generalized 
Wigner crystal \cite{yang}.
It is unclear whether these properties remain in a finite system of coupled 
dots.
The purpose of this paper is to examine general properties of 
a linear chain of spin-polarized dots   
under the condition $N\leq N_{\ell}$.
We find  the following properties.
The chemical potential of the system satisfies,
$(\mu_{N}+\mu_{N_{\ell}+2-N})/2 \approx V+2t$. 
This property of the chemical potential
is reflected in the spacing between the conductance peaks \cite{SBA}.
The electron density structures, 
determined by a quantum Monte Carlo
method, show several types of commensurate/incommensurate
structures
as the number of electrons is varied.
The groundstate electronic
properties of these structures 
cannot be described by mean field theory (Hartree-Fock theory)
except for a few isolated cases.  The reason
for this failure is the existence of many nearly degenerate solutions.

As shown by recent experiments when $N \leq N_{\ell}$ the 
system can be easily spin-polarized  
\cite{Kou2}.  
The Hamiltonian of such a coupled dots
is given by \cite{com1} 
\begin{eqnarray}
H &=& H_t +H_V \nonumber \\
  &=& -t \sum_{i=1}^{N_{\ell}-1}(c^{+}_{i+1}c_{i}+c^{+}_{i}c_{i+1})+2tN \\
  & & +V\sum_{i=1}^{N_{\ell}-1}n_{i}n_{i+1} \nonumber
\end{eqnarray}
The operator $c_{i}(c^{+}_{i})$ 
destroys (creates) an electron in the $i$-th dot.
The term $2tN$ allows to measure single particle energies form zero.

The presence of the end sites at $1$ and $N_{\ell}$ breaks 
particle-hole symmetry.
However, the system can be mapped into a ring of 
$(N_{\ell}+1)$-coupled dots by imposing the periodic boundary
condition $c_{N_{\ell}+2}=c_{1}$ and requiring the electron occupation 
number at the site $N_{\ell}+1$ to be zero: 
\begin{eqnarray}
H&=&-t \sum_{i=1}^{N_{\ell}+1}(c^{+}_{i+1}c_{i}+c^{+}_{i}c_{i+1})\nonumber \\
 & &+V\sum_{i=1}^{N_{\ell}+1}n_{i}n_{i+1}+w_{N_{\ell}+1}n_{N_{\ell}+1}+2tN.
\end{eqnarray}
When the strength of the impurity potential energy, $w_{N_{\ell}+1}$, is large 
and positive the occupation number at the site $N_{\ell}+1$ will be zero.
We rewrite the Hamiltonian using the hole creation and hole
occupation number operators $h_{i}^{+}=c_{i}$ and 
$m_{i}=1-n_{i}$:
\begin{eqnarray} 
H=H'+(V+2t)(2N-N_{\ell}-1),
\end{eqnarray}
where
\begin{eqnarray} 
H'&=&t \sum_{i=1}^{N_{\ell}+1}(h^{+}_{i+1}h_{i}+h^{+}_{i}h_{i+1})\nonumber \\
  & &+V\sum_{i=1}^{N_{\ell}+1}m_{i}m_{i+1}\nonumber \\
  & &+w_{N_{\ell}+1}(1-m_{N_{\ell}+1})+2tN_{h}.
\end{eqnarray}
The quantity  $N_{h}$
is the total number of holes, equal to
$N_{\ell}+1-N$.  From this transformation we find the relation
\begin{eqnarray}
E(N)=E'(N_{\ell}+1-N)+(V+2t)(2N-N_{\ell}-1),
\end{eqnarray}
where $E(N)$ is the groundstate energy of $H$ with $N$ electrons, and 
$E'(N)$ denotes the counterpart of $H'$.  For sufficiently large
$N_{\ell}$ the impurity contributions to the groundstate energies are 
negligible in both $H$ and $H'$, and we may set
\begin{eqnarray}
E'(N)\approx E(N)
\end{eqnarray}
>From this it follows
\begin{eqnarray}
(\mu_{N}+\mu_{N_{\ell}+2-N})/2\approx V+2t.
\end{eqnarray}
For comparison it should be noted that complete electron-hole
symmetry of a ring of $N_{\ell}$ dots leads to
\begin{eqnarray}
(\mu_{N}+\mu_{N_{\ell}+1-N})/2=V+2t. \label{eq:ring}
\end{eqnarray}
Note that the second subscript in Eq.~(\ref{eq:ring}) contains 1
instead of 2. 
In a finite linear chain or in the presence of
disorder this relation
is not valid because electron-hole symmetry is broken.

We have solved Eq.~(1) by a Hartree-Fock method.
The ground state is determined by the following equations 
\begin{equation}
 H_{HF} |k\rangle = \epsilon_k |k\rangle 
\end{equation}
Here the Hartree-Fock Hamiltonian is given by
\begin{eqnarray}
H_{HF}&=&-t\sum_{i=1}^{N_{\ell}-1} (c_i^{\dagger}c_{i+1}+ c_{i+1}^{\dagger}c_i)
+ \sum_{i=1}^{N_{\ell}} X_i c_i^{\dagger}c_i\nonumber\\ 
      & &-V\sum_{i=1}^{N_{\ell}-1}\langle n_{i+1}\rangle \langle n_{i}\rangle.
\end{eqnarray}
where
\begin{displaymath}
 X_i = V(\langle n_{i+1}\rangle + \langle n_{i-1}\rangle) . 
\end{displaymath}
The electron density $\{ \langle n_i\rangle \}$ and the Hartree-Fock
eigenstates $\{ |k\rangle \}$ should be determined self-consistently.
Note that $ \langle n_0\rangle = \langle n_{N_{\ell}+1}\rangle=0$.  
The total energy of $N$-electron system is given in terms of the
Hartree-Fock eigenstates $\{| k\rangle \}$ :
\begin{equation}
 E(N) = \sum_{k=1}^N \epsilon_k - \frac{1}{2}\sum_{k,q=1}^N
 \left( \langle kq |H_V| kq\rangle - \langle kq |H_V| qk\rangle \right),
\end{equation}
where 
\begin{eqnarray*}
 \langle kq |H_V| kq\rangle &=& 2V \sum_{i=1}^{N_{\ell}-1} |a_{i}(k)|^2 
                           |a_{i+1}(q)|^2 , \\
 \langle kq |H_V| qk\rangle &=& 2V \sum_{i=1}^{N_{\ell}-1} 
   a_{i}(k)^{*}a_{i}(q)a_{i+1}(k)^{*}a_{i+1}(q)  . 
\end{eqnarray*} 
The amplitude of the $k$th eigenstate at site $i$ is $a_{i}(k)$.

Fig.~1 displays the lowest HF solutions 
for an odd value of $N_{\ell}=7$.
Since they have nearly degenerate energies  
we must include quantum fluctuations to find the true
groundstate.
Fig.~2 displays
the density profiles obtained using a quantum Monte Carlo
method \cite{Hir}.  
We used typically 500 000 - 1 000 000 Monte Carlo steps to measure
the values and the statistical error is about the size of the symobols.
Quantum Monte Carlo and HF results differ significantly.
The HF theory overestimates the repulsive interactions, and consequently
favors structures with more oscillations. A commensurate
structure is found
at $N=4$.  Generally, when $N_{\ell}$ is odd a commensurate state exists  
for $N=(N_{\ell}+1)/2$.  This type of states have interesting optical properties
\cite{yang}.  
Quantum Monte Carlo results show
that the number of peaks is equal to $N$ when $N\leq N_{\ell}/2$.
The number of minima is equal to $N_{\ell}-N$ for $N > N_{\ell}/2$.

Fig.~3  displays quantum Monte Carlo density profiles for an even value
of $N_{\ell}=14$.  In both cases the following structures are present: 
$(\bullet, \bullet, \circ, \bullet, \bullet, \circ)$,
$(\circ, \circ, \bullet, \circ, \bullet, \circ, \circ, \bullet)$ where
the symbols $\bullet$ and $\circ$ denote relatively large and small occupation 
numbers. HF or classical theory fails to predict these structures.  
Note that when $N_{\ell}/N >1/2$ the occupation number
increases as the site index
moves away from the center and the values of $n_{i}$
are large at $i=1$ and $N_{\ell}$.  
This is because the electrons feel strong mutual repulsion.
The opposite 
is true for $N_{\ell}/N <1/2$.  Note that, in contrast to infinite systems,
the groundstates do not
resemble a liquid state even for $V=1$. Generally when $N_{\ell}$ is even a 
commensurate (periodic) state is absent, since two classically degenerate
state exists. Again we note that
the number of peaks is $N$ when $N\leq N_{\ell}/2$, and the number of 
minima is $N_{\ell}-N$ when $N>N_{\ell}/2$.

The inset in Fig.~4 displays $N$ versus 
$\mu_{N}$ 
for $N_{\ell}=15$ and
$V=4$.  The distance between the $(N+1)$-th and $N$-th peaks is equal to 
$\mu_{N+1}-\mu_{N}$ \cite{SBA}.
The constant charging model would fail to account for
these results since the separations between the peaks are not a constant.
We have tested the accuracy of Eq.~(7) numerically by plotting
$(\mu_{N}+\mu_{N_{\ell}+2-N})/2$ as a function of $N$.  We see in Fig.~5 
that even for small $N_{\ell}$ Eq.~(7) is well satisfied.  

Kouwenhoven {\it et al.}\cite{Kou2} have shown that their experimental
data are better described by assuming presence of some disorder.
We have also carried out a similar calculation in the presence of an impurity,
and find that our result is in qualitative agreement with that of 
Kouwenhoven {\it et al.}\cite{com2}
It would be interesting to test the validity of Eq.~(5) experimentally
in a clean coupled-dot system without impurities.
Strong correlation effects should be more visible in conductance positions 
the larger the ratio $V/t$ is.  A finite chain 
of linearly coupled quantums dots
is well suited for observing both 
commensurate/incommensurate electronic structures.

This work has been supported by  the KOSEF
under grant 961-0207-040-2 and the Ministry of Education under grant
BSRI-96-2444.
M.C.C. has been supported in part by the Ministry of Education under
grant BSRI-96-2448 and by the KOSEF under grant 961-0202-008-2.
K.K. was partially supported by KOSEF-POST-DOC program.
S.R.E.Y. thanks L. P. Kouwenhoven for sending
preprints.

\end{multicols}
 
\begin{figure}
\caption{Occupation numbers, $n_{i}$, as a function of site index, $i$.
The three lowest energy
states of a Hartree-Fock calculation are plotted.  The strength of 
nearest-neighbor electron-electron
interactions, the electron number, and the number of dots are
$V=5$, $N=3$, and $N_{\ell}=7$.  
Here the energies are measured in units of $t$.
The graphs are shifted vertically for the sake of clarity of display.}

\end{figure}
\begin{figure}
\caption{Quantum Monte Carlo results of 
occupation numbers for six different electron numbers $N$.
The parameters are $V=5$ and $N_{\ell}=7$.}
\end{figure}
\begin{figure}
\caption{Quantum Monte Carlo results of 
occupation numbers for nine different electron numbers $N$.
The solid lines are for $V=4$ while
dotted lines are for
$V=1$.  The total site number is $N_{\ell}=14$.}
\end{figure}
\begin{figure}
\caption{Chemical potential, $\mu_{N}$, measured in unit of $t$
is plotted as a function of $N$. 
Results for $(N_{\ell},V)$ equal to (15,4), (14,4), and (14,1) are
represented by circles, inverted triangles, and triangles.  The inset displays
$N$ vesus $\mu_{N}$ for $(N_{\ell},V)=(15,4)$.  The energy 
difference between two adjacent values of $\mu_{N}$ 
corresponds to the spacing between two nearby conductance
peaks.} 
\end{figure}
\begin{figure}
\caption{The quantity $(\mu_{N}+\mu_{N_{\ell}+2-N})/2$, measured in unit of $t$
is plotted as a function of
$N$ to test whether it is approximately equal to a constant $V+2t$.
Results for $(N_{\ell},V)$ equal to (15,4), (14,4), (14,1), (7,4),
and (7,1) are 
represented by circles, inverted triangles, and triangles, diamonds, 
and squares.}
\end{figure}

\begin{references}
\bibitem{K} For recent reviews see L. P. Kouwenhoven and P. L. McEuen, 
in {\it Nano-Science 
and Technology}, editor G. Timp, (to be published);
H. van Houten, C. W. Beenakker, and A. A. M. Staring, in {\it Single 
Charge Tunneling}, edited by H. Grabert and M. H. Devoret 
(Plenum, New York, 1992);
D. v. Averin and 
K.K. Likharev, in {\it Mesoscopic Phenomena in Solids}, edited by B. L. 
Altshuker, P.A. Lee, and R. Webb (Elsevier, Amsterdam, 1991);
U. Merkt, Adv. Solid state Phys. {\bf 30}, 77 (1990),
\bibitem{Kas}M.A.Kastner, Rev. Mod. Phys. {\bf 64}, 849 (1992);
 Phys. Today {\bf 46}, No.1, 24 (1993).
\bibitem{Kou} L. Kouwenhoven, Science {\bf 268}, 1440 (1995).
\bibitem{Wau} F. R. Waugh, M. J. Berry, D. J. Mar, R. M. Westervelt,
K. L. Campman, and A. C. Gossard, Phys. Rev. Lett. {\bf 75}, 705 (1995).
\bibitem{Bli} R. H. Blick, R. J. Haug, J. Weis, D. Pfannkuche, K. v.
Klitzing, and K. Ebert, Phys. Rev.B {\bf 53}, 7899 (1996).
\bibitem{Vaa} N. C. van der Vaart, S. F. Godijn, Y. V. Nazarov, 
C. J. P. M. Harmans, J. E. Mooji, L. W. Molenkamp, and C. T. Foxon,
Phys. Rev. Lett. {\bf 74}, 4702 (1995).
\bibitem{Hof} F. Hofmann, T. Heinzel, D. A. Wharam, J. P. Kotthaus, G. B\"{o}hm,W. Klein, G. Tr\"{a}nkle, and G. Weimann, Phys. Rev.B, {\bf 51}, 13872 (1995).
\bibitem{Mat} K. A. Matveev, L. I. Glazman, and H. U. Baranger,
Phys. Rev.B, {\bf 53}, 1034, (1996).
\bibitem{Gol} J. M. Golden and B. I. Halperin, Phys. Rev. B {\bf 54},
16757 (1996).
\bibitem{Kou2}L. P. Kouwenhoven, F. W. J. Hekking, B. J. van Wess, 
and C. J. P. M. Harmans, Phys. Rev. Lett. {\bf 65}, 361, (1990).
R. J. Haug, J. M. Hong, and K. Y. Lee, Surf. Sci. {\bf 263}, 415 (1992).
\bibitem{yang}  
S.-R. Eric Yang, Phys. Rev. B, to be published (1997), and references therein.
\bibitem{SBA}U. Sivan, R. Berkovits, Y. Aloni, O. Prus, A. Auerbach, and 
G. Ben-Yoseph, Phys. Rev. Lett. {\bf 77}, 1123 (1996);
S.-R. Eric Yang, A. H. MacDonald, and M. D. Johnson,
Phys. Rev. Lett. {\bf 71}, 3194 (1993).
\bibitem{com1}When $N \leq N_l$ and in the presence of a strong magnetic field
electrons will be spin-polarized so that no dot can contain
more than one electron.  In this case the system can be described our
tight binding Hamiltonian (Eq.(1)), and
the effect of the shape of the dot can be included
in the hopping parameter $t$.
\bibitem{Hir}J. E. Hirsch, R. L. Sugar, D. J. Scalapino, and R. Blankenbecler,
Phys. Rev. B {\bf26}, 5033 (1982).
\bibitem{com2}Kouwenhoven {\it et al.} interpretate 
their data using a non-interacting
model, i.e. their system seems to be in the regime $V/t \leq 1$.
In such a case we find,
in agreement with them, that
the total energies come in pairs of two when an impurity is present.

\end{references}
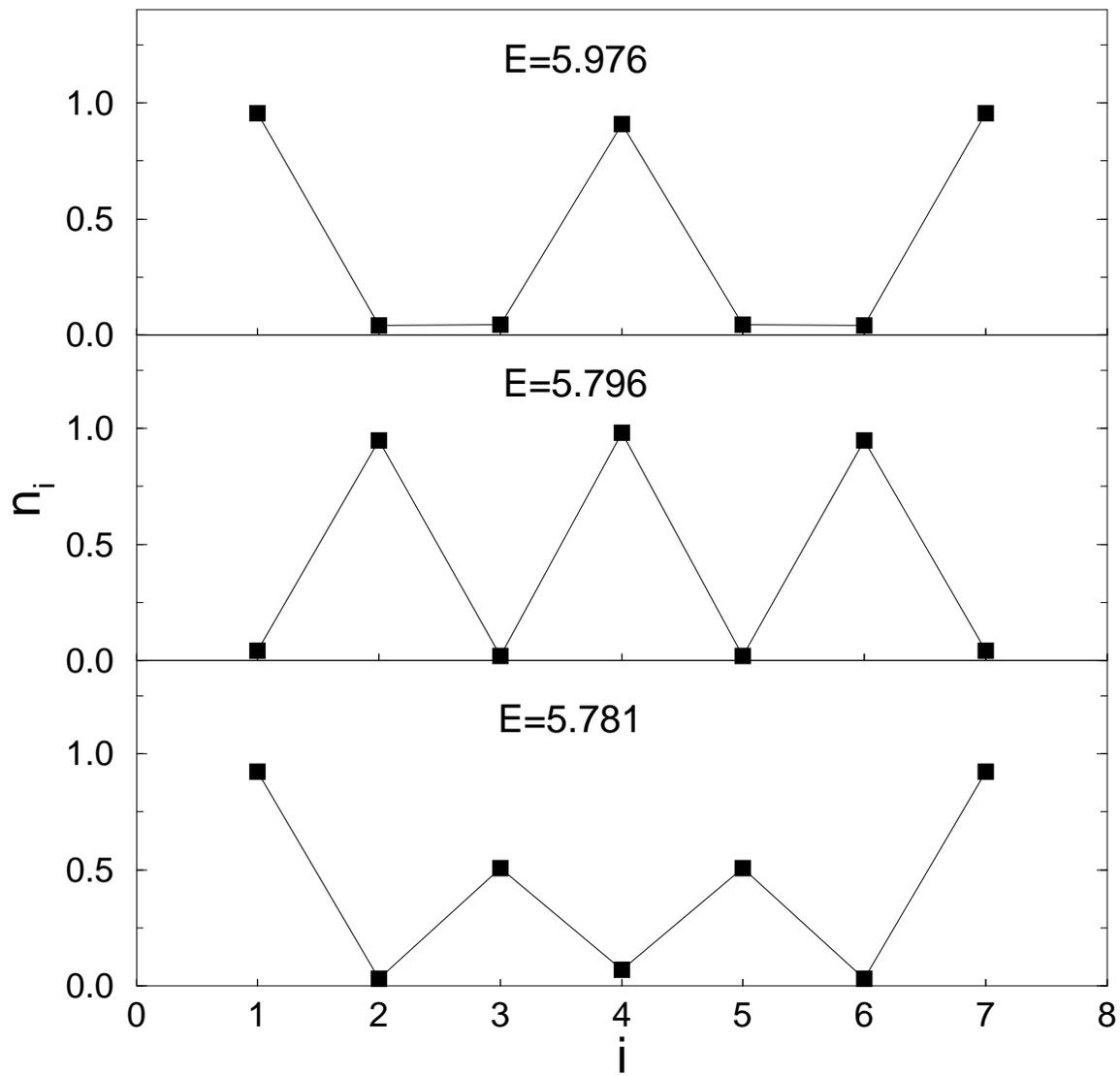
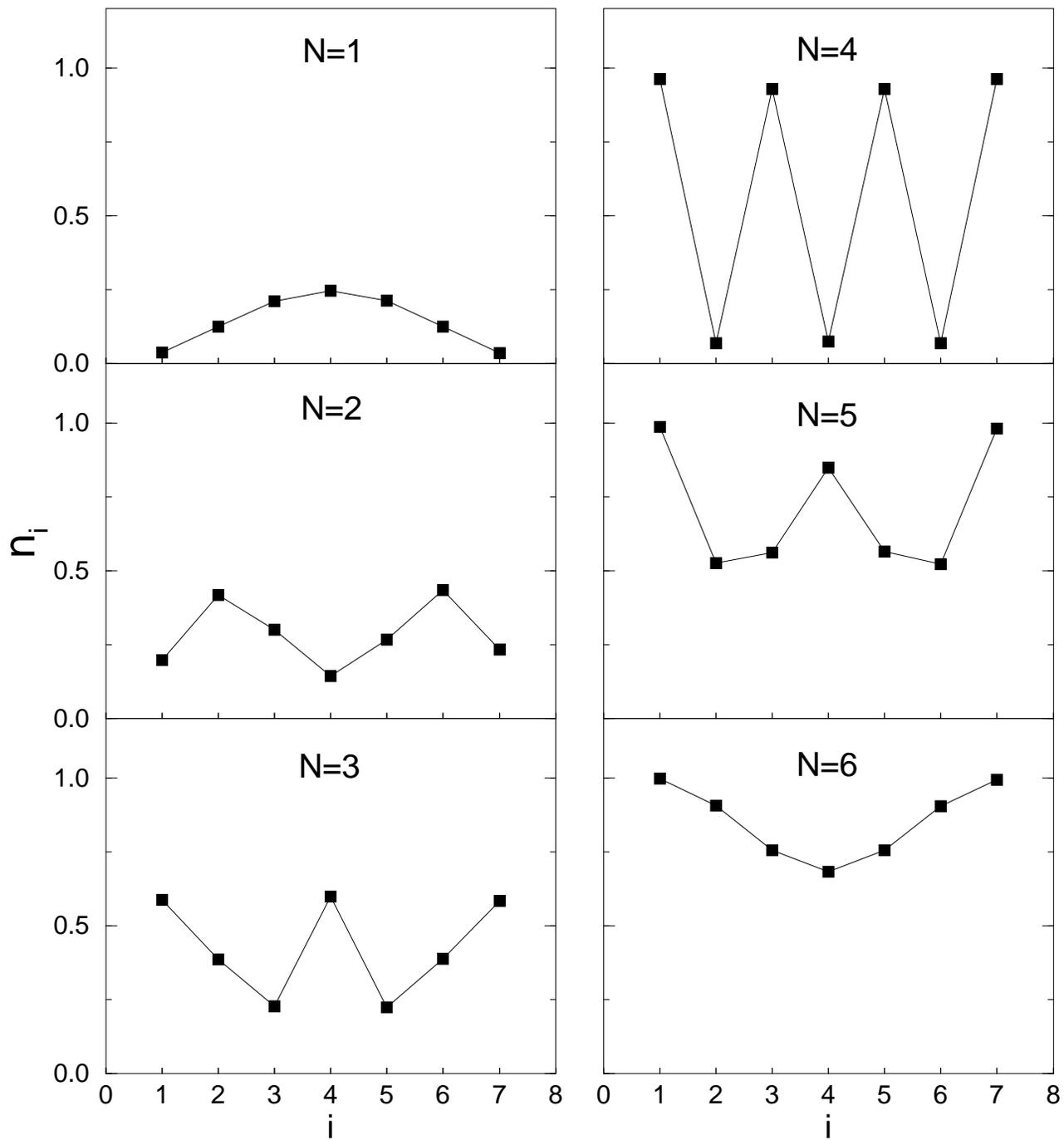
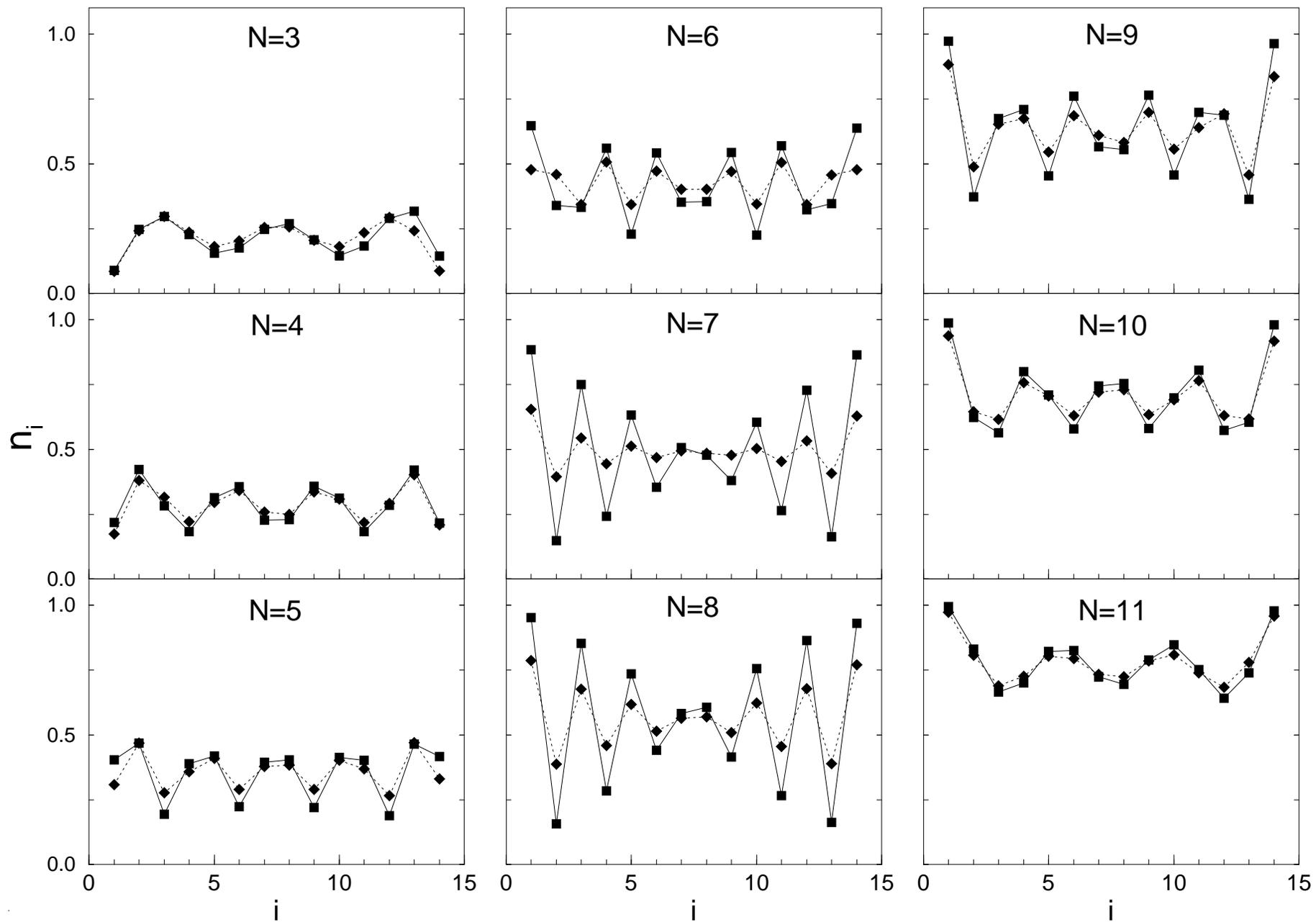
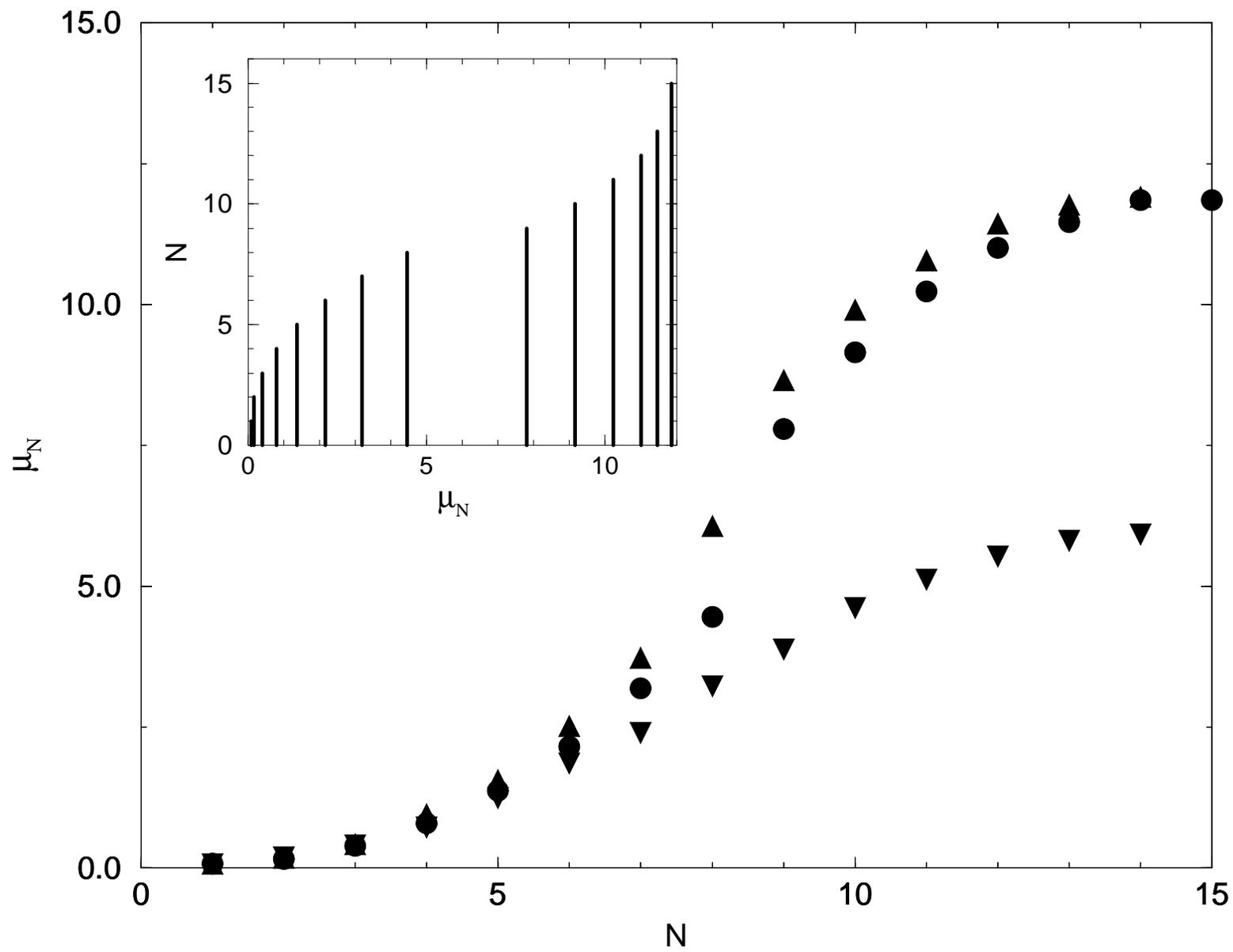
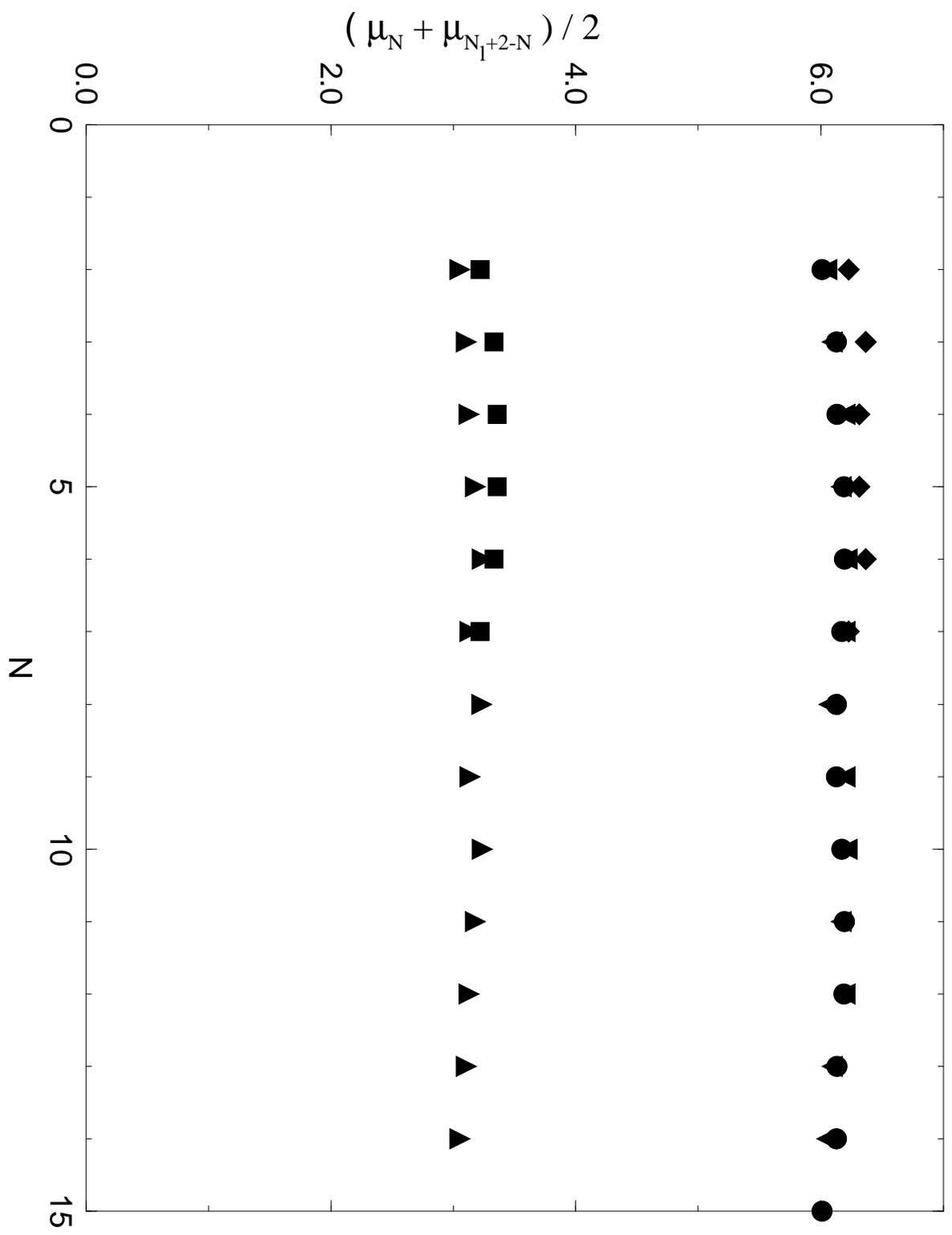
\end{document}